\documentstyle[twocolumn,aps,prl,epsf]{revtex}
\begin{document}
\newcommand{\beq}{\begin{equation}}
\newcommand{\eeq}{\end{equation}}

\title{Dependence of folding rates on protein length}

\author{Mai Suan Li$^1$, D. K. Klimov$^2$ and D. 
Thirumalai$^2$}

\address{$^1$Institute of Physics, Polish Academy of Sciences,
Al. Lotnikow 32/46, 02-668 Warsaw, Poland\\
$^2$Institute for Physical
Science and Technology, University of Maryland, College Park, MD 20742 USA}

\address{
\centering{
\medskip\em
{}~\\
\begin{minipage}{14cm}
Using three-dimensional Go lattice  models with side chains for
proteins, 
we investigate the dependence of 
folding times on protein length. 
In agreement with previous theoretical predictions, we find that the folding
time $\tau_F$ grows as a power law with the chain length, i.e., 
$\tau_F \sim N^\lambda$, where 
$\lambda \approx 3.6$ for the 
Go model, in which all native interactions (i.e., between all 
side chains and  backbone atoms) are uniform.   If the interactions
between side chains are given by pairwise statistical potentials,
which introduce heterogeneity in the contact energies, then
the power law fits yield large 
$\lambda$ values that typically signifies a crossover to an
underlying activated process. Accordingly,   
the dependence of $\tau_F$ on $N$ 
is best described using $\tau_F \sim
e^{\sqrt{N}}$.    
The study also shows that the incorporation of side chains considerably
slows down folding by introducing energetic and topological
frustration.  
{}~\\
{}~
\end{minipage}
}}

\maketitle

\section{Introduction}

Protein folding mechanisms depend not only on the architecture of the
native state, but on the external conditions (pH, salt concentration,
temperature, and molecular environment). Several recent studies  have argued
that the folding rates (presumably, under the conditions of neutral
pH, zero salt concentration, room temperature, and the absence of
molecular crowding) is determined solely by the
architecture of the native structure \cite{Plaxco}. 
Although the native  topology
does constrain the ensemble of transition states (the folding nuclei
have to be topology preserving \cite{GuoThirum95}), 
other factors, such as protein size
and the native state stability, also play a role in determining folding
rates and mechanisms. For instance, 
a direct correlation between
folding rates and stability has been noted by Clarke and
coworkers \cite{Clarke99Structure}. They showed that for five proteins, all
with immunoglobulin-like fold, the folding rates $k_F$ correlate well with
the native state stability. On the other hand, 
there is a poor correlation between $k_F$
and the relative contact order \cite{Plaxco}, 
which quantifies the balance of local vs non-local native interactions. 
Improved correlation may still be expected for the proteins with
$\alpha$-helical or $\alpha/\beta$ architecture \cite{Oliverberg01JMB}.

Although the importance of native state stability in determining $k_F$
has been demonstrated, limited experimental data have been used
to argue that the length of proteins
(i.e., the number of amino acids $N$) should not affect $k_F$
\cite{Plaxco}. From the
polymer physics perspective, this is somewhat surprising, because the
relaxation rates even for ideal  polymer 
chains depend on $N$. For example, the largest relaxation time in a
Rouse chain scales as $N^2$ \cite{deGennesbook}. 
Because the size range of single
domain proteins is limited (typically less than about 200 residues),
the dependence of $k_F$ on $N$ cannot be sharply demonstrated. In
proteins other factors, such as amino acid sequence and the nature of local and
non-local interactions in the native state, could be more
dominant. Nevertheless, the mere fact that proteins are polymers
implies that $N$ should play some role in determining the folding
rates \cite{Takada}. 

The dependence of $k_F$ on $N$ has been investigated in a number of
theoretical studies 
\cite{Thirumalai,Wolynes1,Finkelstein,Wolynes2,Shakhnovich,Cieplak,Ball}. 
Several folding scenarios emerge depending on the
characteristic folding temperatures, namely, the collapse temperature,
$T_\theta$, the folding transition temperature, $T_F$, and the
glass transition temperature, $T_g$ \cite{Thirumalai,Bryn95}. 
For optimal folding, for which $T_\theta
\approx T_F$, Thirumalai has predicted that the folding time $\tau_F$
should scale with $N$ as \cite{Thirumalai} 
\begin{equation}
\tau_F \sim N^\lambda. 
\label{taulambda}
\end{equation}
The dimensionality dependent 
exponent $\lambda$ for two-state folders is expected to be
between 3.8 and 4.2 \cite{Thirumalai}. 
Simulation studies using Go lattice models (LMs) without side chains suggest
a smaller value of about 3 \cite{Shakhnovich,Cieplak}. 
These numerical studies are in broad agreement with the
theoretical predictions.  The heteropolymer nature of protein-like
models could make  $\lambda$ temperature dependent. For two-state
optimized folders there is a relatively broad range of temperatures,
where $\tau_F$ remains relatively insensitive to $T$ \cite{Abkevich98}.  
The $N$ dependence of
$\tau_F$ outside this range may not obey Eq. (\ref{taulambda}) or
$\lambda$ may be different. 

All of the numerical studies mentioned above have been done using
LMs, in which each residue is represented by a bead
confined  to the vertices of an appropriate (usually cubic)
lattice. Side chain packing effects, which are crucial in the 
folding process, cannot be considered in this class of LMs.   
A simple way to include these in the context of LMs 
is to attach additional bead to each $\alpha$-carbon atom in a
sequence \cite{Bromberg,Klimov}. Thus, an amino acid consists now 
of two beads, one
representing a backbone (BB) and the other - a side chain (SC). In 
this polypeptide model there are  2N beads. If an
appropriate heterogeneous potential between side chains is included,
then the cooperative transition reminiscent of folding can be
reproduced \cite{KlimThirum02JCC}. 
Thermodynamics and kinetics of lattice models with side
chains (LMSC) have been
recently reported \cite{Klimov,KlimThirum01Proteins,Li}. 

In this paper we examine the effect of rotamer
degrees of freedom on the exponent $\lambda$ (Eq. (\ref{taulambda}))
using Go-like models \cite{Go}. 
In these   highly simplified models, which have
received considerable attention in recent years, only interactions
present in the native state are considered. Non-native interactions,
which can play an important role in the thermodynamics and kinetics of
folding \cite{KlimThirum01Proteins,Li,MSLi}, 
are ignored. Nevertheless, several
studies have showed that the Go models provide a reasonable caricature
of certain aspects of folding 
\cite{KlimThirum01Proteins,Onuchic00JMB,KlimThirum00PNASa}. 
For this model system, we find that
the exponent $\lambda$ is altered by rotamer degrees of freedom. More
importantly, $\lambda$ depends on the {\em details of interactions} and, in
this sense, is non-universal. As a technical byproduct of our investigation
we show that robust results for 
$\lambda$ are obtained only for those Monte Carlo move sets, which 
are  effectively ergodic.

\section{Methods}

\noindent
{\bf Model:} In the LMSC 
the energy of a conformation is  \cite{Klimov}
\begin{eqnarray}
E \; = \;  \epsilon _{bb} \sum_{i=1,j>i+1}^{N} \, \delta _{r_{ij}^{bb},a} 
+ \epsilon _{bs} \sum_{i=1,j\neq i}^{N} \, \delta _{r_{ij}^{bs},a} 
+ \nonumber\\ \epsilon _{ss} \sum_{i=1,j>i}^{N} \, \delta _{r_{ij}^{ss},a} \; ,
\label{eqn2}
\end{eqnarray}
where $\epsilon _{bb}, \epsilon _{bs}$ and $\epsilon _{ss}$ are BB-BB,
BB-SC and SC-SC contact energies.
$r_{ij}^{bb}, r_{ij}^{bs}$ and $r_{ij}^{ss}$ are 
the distances
between the $i^{th}$ and $j^{th}$ residues for the  BB-BB,
BB-SC and SC-SC pairs, respectively.
Each lattice site can only be occupied by a single bead (BB or SC) so that
the self-avoidance condition is satisfied. 

We consider two versions of the Go model. In the model GM1,   
$\epsilon _{bb}, \epsilon _{bs}$ 
and $\epsilon _{ss}$ are chosen to be -1 for native contacts 
and 0 for non-native ones.
In GM2, $\epsilon _{bb}=\epsilon _{bs}=-0.2$ and  the values of
$\epsilon_{ss}$, which depend on the nature of amino acids, are given
by Betancourt-Thirumalai statistical interaction potentials 
\cite{Betancourt99}. Thus, GM2 incorporates diversity in the
interaction energies that is known to be important in the design of
foldable sequences. 
The fraction of hydrophobic residues
in a sequence is approximately 0.5 as in wild-type proteins. By
setting all of non-native 
contact interactions to  3.0 (this value is larger than any of 
Betancourt-Thirumalai couplings \cite{Betancourt99}), we ensure that
non-native interactions do not contribute to 
folding. Thus, for all practical purposes both models exhibit Go-like
characteristics. 

The maximum length of sequences $N$ examined in our work is
40. Investigation of scaling behavior of LMSC beyond this limit 
is computationally expensive. However,
scaling trends may be reasonably 
established for LMs without side chains even for
$N\le 40$ (see Fig. (2) in \cite{Shakhnovich}). 
Therefore, we are  fairly confident that considering LMSC
with $N\le 40$ would not hamper our ability to analyze scaling
of folding times  with $N$. 

\noindent
{\bf Sequences:} Protein-like sequences were obtained using the standard
$Z$-score optimization  or by minimizing the energy of
the native state \cite{Z-score,Shakhnovich94PRL}. 
$Z$-score optimization  is based on 
Monte Carlo simulations in sequence space aimed at 
minimizing $Z=(E_0-E_{ms})/\delta$, where $E_0$ and $E_{ms}$ are
the energy of a sequence in the target conformation and the average energy of
misfolded structures, respectively, and $\delta$ is the dispersion in
the native contact energies.  We took 
$E_{ms}=c<B>$, where $c$ is average number of 
nearest neighbor contacts in the manifold of 
misfolded structures, and $<B>$ is the 
average contact energy for a given sequence. The Monte Carlo
simulations in sequence space were done using simulated annealing
protocol by generating 20 independent trajectories for each sequence. The
optimized sequence is the one with the lowest $Z$-score. For each $N$,
the target conformations for GM1 and GM2 are identical, so the  effect
of the "realistic" SC interactions can be directly addressed. 

Despite its simplicity $Z$-score and energy optimizations
\cite{Z-score,Shakhnovich94PRL} 
have been proven to be effective
techniques for generating designed foldable sequence with $N\lesssim
100$ \cite{Shakhnovich}. Other, more elaborate, technical methods for
designing lattice protein-like sequences have been proposed
\cite{Betancourt}. For the scope of our paper these methods are not
relevant, because we are only interested in generating foldable sequences
spanning a reasonable range of $N$.

\noindent
{\bf Move sets in Monte Carlo simulations:} To assess the efficiency
of the Monte Carlo (MC) 
simulations and to check the robustness of the results we
used four distinct move sets (Fig. (1)). Move set 
MS1 involves only single corner (and also tail) monomer moves.
In addition to single flips,  the standard move set (SMS) also contains 
the crankshaft motion \cite{Hilhorst}. 
Fig. (1) shows the moves in the set MS2, which includes SMS and additional
two-monomer moves (not crankshaft ones). Move set MS3 is implemented
as described in \cite{Betancourt}. The validity of MS3 has been verified 
for short sequences without side chains by comparing MC results with
those obtained by full enumeration of lattice conformations, which is
tantamount to performing ensemble average. Thus,  for MS3 ergodicity 
has been established and implementation of detailed
balance condition has been also discussed \cite{Betancourt}. 
We have found that due to its  flexibility 
MS3 {\em is far  more efficient than others}. The purpose of using
different MC move sets is to ensure the robustness of our results. 
In our study 
one MC step consists of $N$  MC moves, i.e., on an
average each bead in a sequence is attempted to move once during one MC step.

\noindent   
{\bf Computation of folding times and temperatures:} For each 
sequence and temperature we computed the distribution of first
passage times $\tau_{1i}$, where $\tau_{1i}$ is the number of MC steps
needed to reach  the native state starting from the 
unfolded state $i$. The structure is considered folded, if the overlap
function $\chi = 0$ \cite{Klimov}. The folding time $\tau_F = \frac1M
\sum_{i=1}^M \tau_{1i}$, where $M=100$ is the number of individual
trajectories. The folding time to reach the native
backbone, $\tau_F^{bb}$, has been calculated in a similar way. 

In our study, folding times $\tau_F$ have been obtained at the
temperature of fastest folding $T_{min}$ located by scanning a
temperature range for each sequence. If the folding transition
temperature $T_F$ is identified with the maximum in the fluctuations
of overlap function \cite{Klimov}, 
then we expect $T_F\approx T_{min}$. 
This conclusion, which naturally follows from the dual requirement  of the
stability of protein native state and its kinetic accessibility, is 
illustrated in Fig. (2) for two-state LMSC sequence with $N=15$. The
observation that $T_F\approx T_{min}$ serves as a convenient
operational condition to pinpoint the temperature of folding
simulations without the need of expensive equilibrium simulations. 
Furthermore, it has been shown \cite{Abkevich98} 
that for highly optimized
sequences there is a large plateau in the temperature dependence of
$\tau_F$. This flat temperature  range typically includes $T_{min}$, $T_F$, and
collapse temperature $T_{\theta}$. We expect that scaling
behavior would be similar for these  temperatures. 

In a previous study \cite{Cieplak} 
that has examined the $N$ dependence of folding
times, $T_F$ has  been defined as the temperature, at which the
probability of occupancy of the microscopic native conformation is
0.5. Such a highly restrictive definition of $T_F$ is not physically
meaningful. It is realized that the
fluctuations in the overlap function or the equilibrium  fraction of
native contacts are more appropriate quantities 
for defining $T_F$ \cite{Klimov,Li,Shea}. The physically relevant
definition, which also coincides with the experimental ones, is based
on the notion of the native state as a {\em collection} of structurally
similar conformations belonging
to the native basin of attraction (NBA).

\section{Results}

We have monitored the length dependence of the folding times $\tau_F$
(which register folding of the entire native structure) as well as
$\tau_F^{bb}$, which is the average first passage time to the folded
backbone. 
The characteristic U-shape for the temperature dependence of
$\tau_F$ noted for optimized sequences \cite{Shakhnovich} is also found for
$\tau_F^{bb}$ (Figs. (3)). 
Over the range of temperatures, where $\tau_F$ remains
roughly constant, $\tau_F \approx \tau_F^{bb}$. However, this is not
the case at low and high temperatures (see below). 
The extent of the plateau  region
in $T$ is  larger for GM1 (Fig. (3)). 
Narrow shape of the temperature dependence of
$\tau_F$ for GM2 resembles that computed for the random sequence without
side chains (Fig. (3) in \cite{Shakhnovich}).   
In the rest of the paper we focus on the
variations of the folding times at $T_{min}\approx T_F$.

The $N$-dependence of folding times obtained by four types of
MC move sets for GM1 at $T=T_{min}$ is presented in Fig. (4). 
The number of targets used for
$N=9$, 15, 18, 24, 28, 32, and 40  are 100, 50, 50, 20, 17, 15,  and 15,
respectively. For MS1 the calculations were performed only up to $N = 32$.
Ergodic move sets (i.e., SMS, MS2, and MS3), which efficiently sample
the conformational space, were used for $N=40$.

It is well known that MS1 based on single monomer corner and tail flips is not
ergodic \cite{Hilhorst}. Consequently,
the folding rates obtained using MS1 are not reliable. For GM1
we computed $\lambda = 3.7 \pm 0.3, 3.6 \pm 0.2$ and $3.6 \pm 0.2$
for MS2, MS3 and SMS, respectively (Fig. (4)). The 
power law also holds for folding of the backbone with 
$\lambda_{bb} \approx \lambda$ at the simulation temperatures
$T_{min}$. 
Interestingly, exponents $\lambda$ and $\lambda_{bb}$ obtained by
the three ergodic move sets are almost the same, which establishes the
robustness of our results. 
Because the scaling exponent $\lambda$ for Go LMSC models 
is higher than for those without SCs
\cite{Shakhnovich,Cieplak}, we assume that dense packing of 
side chain creates additional barriers to folding.
We also note that $\lambda$ for GM1 sequences  is in the 
range proposed by Thirumalai for fast folding sequences \cite{Thirumalai}.

Side chains alter $\lambda$ values for GM2 with 
realistic interactions. It is still possible to fit the folding times
obtained for GM2  using  a power law with the large exponent 
$\lambda \approx 6.5 \pm 0.4$, which is similar to 
that reported  for random sequences without side chains \cite{Shakhnovich}. 
Because the same target
conformations for both GM1 and GM2 simulations have been used, the large 
difference
in exponents is due to the diversity of interactions. 
Thus, our study shows that, even though GM2
sequences exhibit two-state folding, heterogeneity in the interactions between
side chains  (as in GM2) adds 
roughness  to the underlying energy landscape. 
Furthermore, side chain packing also
introduces enhanced topological frustration compared to LM without
side chains. As a consequence scaling behavior of those sequences resembles
that of random sequences without side chains \cite{Shakhnovich}. In fact,
for random sequences without side chains Gutin {\em et al} had not ruled
out the possibility of exponential scaling \cite{Shakhnovich}. 

From the physical viewpoint large
$\lambda$ values are indicative of an activated process with a free
energy barrier scaling slower than  $N$ \cite{Wolynes2}. 
In particular, using the proposal
that the activation barrier scales as $\sqrt{N}$ \cite{Thirumalai},  
we find that $\tau_F$ for GM2
can be fit by $\tau_F \sim e^{\beta
\sqrt{N}}$ (see Fig. (5)).  
Recent analysis of experimental data also suggests that the
barrier height scales as $N^\alpha$ with $\alpha = 0.607 \pm 0.179$ 
\cite{Takada}, which is consistent with $\sqrt{N}$ \cite{Thirumalai} 
or $N^{2/3}$ \cite{Wolynes2} scaling.

Although the scaling exponents are the same for all three ergodic
move sets, the folding times vary. 
The dependence of $\tau_F^{SMS}/\tau_F^{MS2}$  
and  $\tau_F^{SMS}/\tau_F^{MS3}$ 
on $N$ for GM1 shows that the folding times
obtained by the standard move set SMS \cite{Li} are about twice  as
long as  those for MS2 and MS3. Because 
$\tau_F^{SMS}/\tau_F^{MS3} >\tau_F^{SMS}/\tau_F^{MS2}$, we conclude 
that the MS3 dynamics is the most efficient for folding in LMs.
This is not unexpected, because  MS3 incorporates flexible choice of 
multimeric  moves, which efficiently sample
local conformational space. 

Both GM1 and GM2 show that there are no significant differences in the
time scales for backbone and side chain folding in the plateau
temperature range, i.e. $\tau_F^{bb}/\tau_F \sim O(1)$ 
(Fig. (3)). However, outside this temperature range $\tau_F^{bb}$
starts to deviate significantly from $\tau_F$. Of particular interest
are the temperatures $T < T_F$, where NBA is populated. This shows
that at low temperatures ($T/T_F$ is relatively small) the backbone
ordering occurs considerably faster than the folding of side
chains. The rate determining step is associated  side chain ordering,
which might involve transitions over barriers of varying
heights. Thus, there may be a relatively narrow temperature window
(for example, $0.9 \lesssim T/T_F \lesssim 1.2$ in Fig. (2)), in
which $\tau_F^{bb} \approx \tau_F$.

\section{Conclusions}

We have studied the scaling properties of Go lattice sequences with
SCs using four different types of Monte Carlo moves. The exponents in the  
power laws describing the scaling of folding times with sequence
length $N$ 
are sensitive to the ergodicity of the move sets and
interaction details \cite{Shakhnovich}. 
The move set MS3, which is based on  flexible selection of 
multimeric moves, is found to be the 
most efficient for studying folding in LMs. 
Strong dependence of folding times for LMSC on sequence length is
attributed to side chain packing. The
presence of side chains interacting via diverse potentials gives rise
to intrinsic roughness in the underlying energy landscape.

\acknowledgments 
It is a
pleasure to dedicate this paper to John Tully on the occasion of his 60th
birthday. Fruitful discussions with M. Betancourt and R. Dima are gratefully 
acknowledged.  This work was supported by KBN (Grant No. 2P03B-146-18) and the
grant from the National Science Foundation (CHE02-09340).

\newpage


\begin{figure}
\epsfxsize=3.2in
\centerline{\epsffile{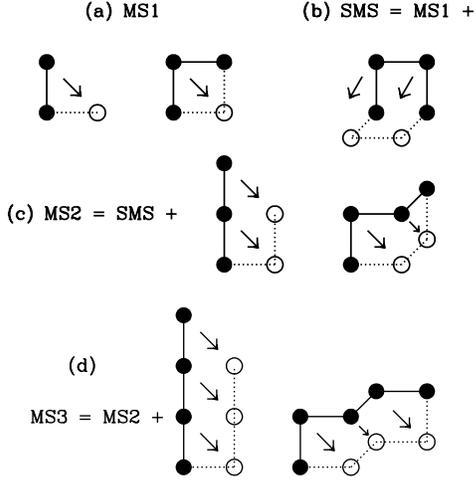}}
\vspace{0.2in}
\caption{ 
MC move sets examined in this study: (a)
MS1 is based on single monomer  corner flips (including tail ones);
(b) SMS incorporates MS1 and the crankshaft moves; (c) MS2 involves SMS
and additional two-monomer moves;
(d) MS3 contains MS2 and three-monomer moves [28].}
\end{figure}

\begin{figure}
\epsfxsize=3.2in
\centerline{\epsffile{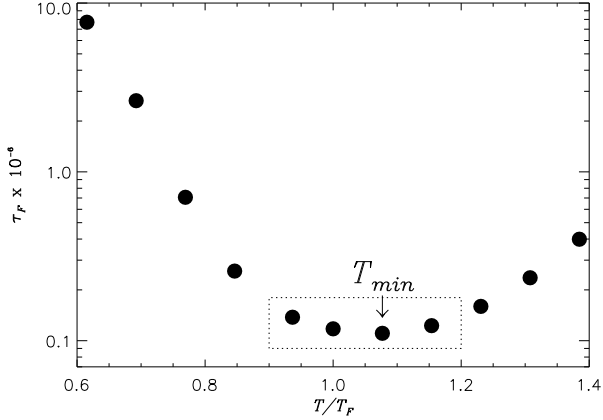}}
\vspace{0.2in}
\caption{
The characteristic U-shape dependence of the folding
time $\tau_F$ on temperature measured in the units of the folding
temperature $T_F$ for the LMSC sequence A ($N=15$) studied in
[17].  $T_F$ is computed using multiple histogram method as
the temperature, at which the fluctuations in the overlap function 
reach maximum. The temperature, at which $\tau_F$ is minimum, 
is  $T_{min}=1.07T_F$. An observation that 
$T_F \approx T_{min}$ may be used for 
crude estimation of $T_F$ for a large dataset of sequences. 
Almost  flat region in the vicinity of $T_F$ is indicated by  a
dotted box [15] (see also Fig. (3)). }  
\end{figure}

\begin{figure}
\epsfxsize=3.2in
\centerline{\epsffile{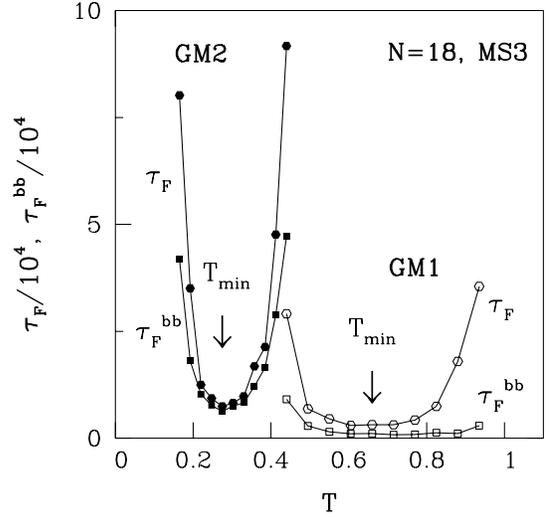}}
\vspace{0.2in}
\caption{
The temperature dependence of $\tau_F$ and $\tau_F^{bb}$ for 
$N=18$ GM1 and GM2 sequences, both of which share 
the same native conformation. Data are obtained using MS3. 
The arrows indicate $T_{min}$, the temperature at which $\tau_F$ is
minimal. The GM2 sequences have a narrower plateau region than
GM1. Therefore, GM2 sequences are not as well optimized as GM1.}
\end{figure}

\newpage

\begin{figure}
\epsfxsize=3.2in
\centerline{\epsffile{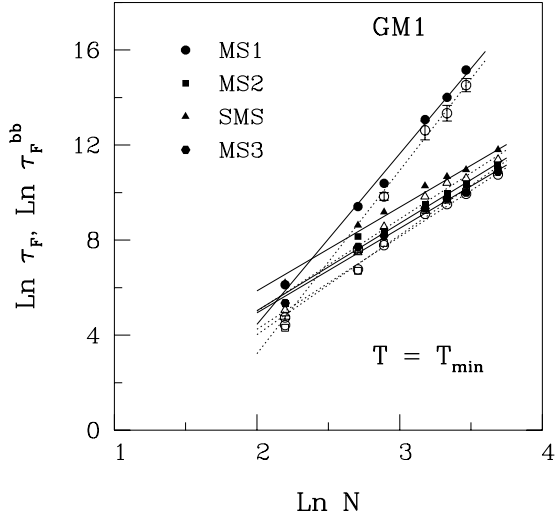}}
\vspace{0.2in}
\caption{
Scaling of $\tau_F$ and  $\tau_F^{bb}$
for GM1 at $T=T_{min}$. The results are shown  for all 
four types of MC move sets. The solid symbols
indicate folding times for the entire sequence $\tau_F$, 
the open ones represent backbone folding times $\tau_F^{bb}$.
Straight solid and dotted lines are corresponding power law
fits. Scalings of $\tau_F$ and  $\tau_F^{bb}$ are virtually
identical, i..e, $\lambda \approx \lambda_{bb}$. The scaling exponents for MS1
move set substantially differ from the others 
that reflects  its inherent lack of ergodicity. 
The results are averaged over  100, 50, 50, 20, 17, 15  and 15 target
conformations for $N$=9, 15, 18, 24, 28 , 32 and 40, respectively.}
\end{figure}

\begin{figure}
\epsfxsize=3.2in
\centerline{\epsffile{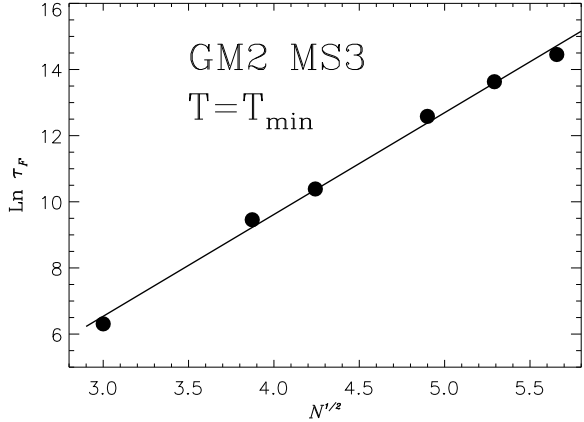}}
\vspace{0.2in}
\caption{
Scaling of $\tau_F$  with $N$ for GM2 sequences
computed at $T=T_{min}$ using MS3. 
The solid line represents the  exponential fit
$e^{\beta \sqrt{N}}$ with $\beta = 3.1\pm 0.5$.  Exponential fit  provides a 
physically sound interpretation for such scaling behavior 
based on barrier crossing.}
\end{figure} 

\end{document}